\documentclass[twocolumn, amsmath,amssymb, superscriptaddress]{revtex4}

\usepackage{dcolumn}% Align table columns on decimal point
\usepackage{bm}% bold math
\usepackage{amsmath}
\usepackage{amsfonts}
\usepackage{graphics}
\usepackage[pdftex]{graphicx}
\usepackage{times}
\usepackage{setspace}
\usepackage{verbatim}
\usepackage{color}   % {\color{red} TEXT }

\begin{document}

\title{The case for caution in predicting scientists' future impact}
\author{Orion Penner}
\affiliation{Laboratory of Innovation Management and Economics, IMT Lucca Institute for Advanced Studies,  55100 Lucca, Italy}
    \author{Raj K. Pan}
\affiliation{Department of Biomedical Engineering and Computational
  Science, Aalto University School of Science, P.O. Box 12200, FI-00076,
  Finland} 
  \author{Alexander M. Petersen}
  \affiliation{Laboratory for the Analysis of Complex Economic Systems,  IMT Lucca Institute for Advanced Studies,  55100 Lucca, Italy}
\author{Santo Fortunato}
\affiliation{Department of Biomedical Engineering and Computational
  Science, Aalto University School of Science, P.O. Box 12200, FI-00076,
  Finland}

%\date{\today}

\begin{abstract}
We stress-test the career predictability model  proposed by Acuna et al. [Nature 489, 201-2 2012] by applying their model to a longitudinal career data set of 100 Assistant professors in physics, two from each of the top 50 physics departments in the US. The Acuna model claims to predict $h(t+\Delta t)$, a scientist's $h$-index  $\Delta t$ into the future, using a linear combination of 5 cumulative career measures taken at career age $t$. Here we  investigate how the ``predictability'' depends on the aggregation of career data across multiple age cohorts. We confirm that the Acuna model does a respectable job of predicting $h(t+\Delta t)$ up to roughly 6 years into the future when aggregating all age cohorts together. However, when calculated using subsets of specific age cohorts (e.g. using data for only $t=3$),  we find that the model's predictive power significantly decreases, especially when applied to early career years. For young careers, the model does a much worse job of predicting future impact, and hence, exposes a serious limitation. The limitation is particularly concerning as early career decisions make up a significant portion, if not the majority, of cases where quantitative approaches are likely to be applied. 
\end{abstract} 
% 1000 words

\maketitle 
\footnotetext[1]{ Published in {\it Physics Today} {\bf 66}, 8--9 (2013). 
\\ Send correspondence to:\\ \text{petersen.xander@gmail.com or santo.fortunato@aalto.fi}}
Any scientist pursuing a research career these days is acutely aware of the
increasingly central role metrics play in measuring scientific impact. From
papers to people, the quality of almost everything is being measured by
citations. One area in which metrics are starting to cause a shift is in
the scientific career evaluation  process. From a purely economic point of
view, a tenure track hire is a million dollar bet on a young scientist's
future success, so it is easy to see why metrics and models capable of
predicting future success are very attractive to decision makers, but it
also highlights that this ``genie'' is unlikely to be put back in its
bottle. 

If metrics are going to be integrated into the career advancement process
they must be better tested and many specific questions need to be
investigated. For example, what aspects of a career are {\it actually}
predictable? What ingredients are required for a model to be robust? How
often is a given model's prediction wrong, and what impact does that have
on the careers of scientists, especially young ones that are already
burdened by risk \cite{GrowthCareers}? Without a proper
  understanding of the above questions, any uncritical use of quantitative
  indicators can do real harm to scientists
%  , who are shown the door based on bogus evaluations, 
  and to the endeavor of science as a
whole.

\begin{figure}[t]
\centering{\includegraphics[width=0.5\textwidth]{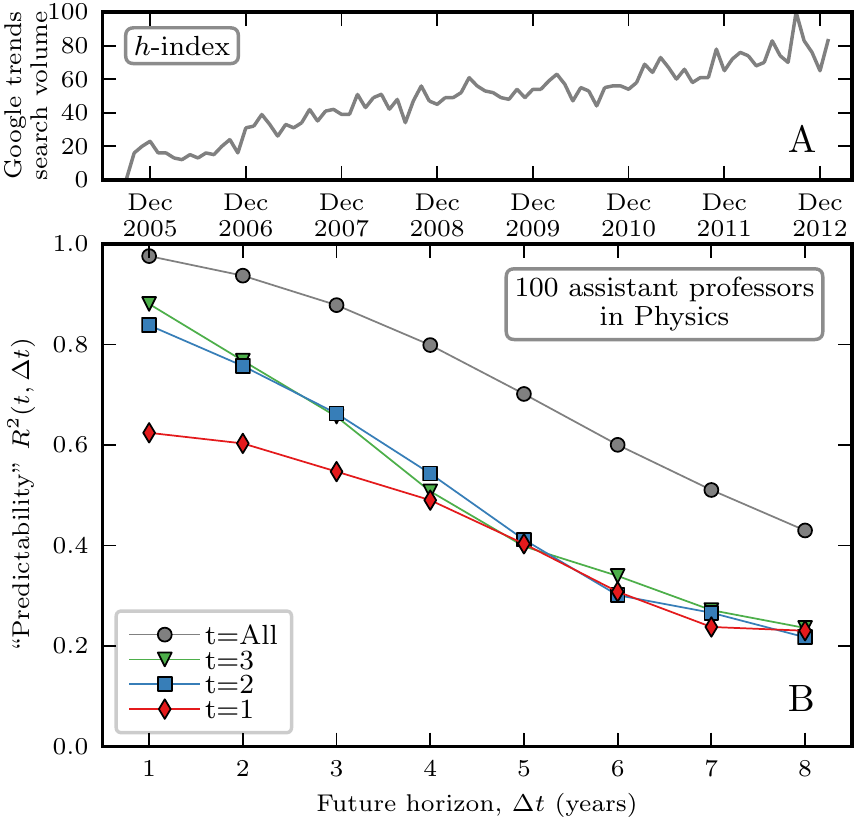}}
\caption{\label{f1} (A) Google search volume, normalized to \% peak value, is a proxy for the interest in the ``h-index''. (B) The ``predictive power'' of the Acuna model \cite{Acuna} decreases significantly when early career age cohorts (years since first publication $t=1,2,3$) are analyzed separately.}
\end{figure}

The introduction of the {\it h}-index~\cite{hirsch2005} in late 2005 was a
significant milestone in the use of metrics in  career evaluation. Figure
1(A) shows that the popularity of the $h$-index has consistently increased since its
introduction. Now it stands as the most popular quantitative measure for a
researcher's productivity and impact. In fact, it is already being
used to evaluate scientists, e.g., a modified version of
the {\it h}-index has been integrated into the Italian national tenure
competition~\cite{ANVUR}.

Of course future impact, rather than past accomplishments, is really what's
at the heart of most career appraisal decisions in science, e.g., tenure,
grants, fellowships, prizes, etc. So how predictable is an individual's
future {\it h}-index? Previously, it has been indicated that the
{\it h}-index is better than other indicators in predicting future
scientific achievements~\cite{hirsch2008}. A more recent publication by Acuna
{\it et.  al.}~\cite{Acuna} presents a model that predicts an individual's
future {\it h}-index using a linear combination of five other metrics. In
its technical details this work is notable because it is one of the
first to integrate several metrics into a prediction. However it is
probably more noteworthy in its non-technical aspects given the
high profile forum in which it was published and the authors' suggestion
that it can be used in decision making, going as far as to provide
an online future {\it h}-index calculator.

\bigskip
In the Acuna {\it et. al.} model, an individual's future {\it h}-index, $h(t+\Delta t)$, is modeled as a linear combination of his/her (i)  current {\it h}-index $h(t)$, (ii) square root of number of publications $\sqrt{N}$, (iii) number of years since first publication $t$, (iv) number of publications in high impact journals $q$, and (v) number of distinct journals $j$.  By incorporating several key aspects of academic publishing their multiple regression model appears quite promising. However further investigation highlights the care that needs to be taken in developing models of future impact.

To illustrate the difficulties of predicting future success we applied the Acuna model to a longitudinal career data set of 100 Assistant professors in physics, two from each of the top 50 physics departments in the US (see \cite{GrowthCareers} for further description of this dataset). Figure 1(B) shows the coefficient of determination $R^{2}$ for the regression model $\Delta t$ years into the future using data available at ``career age'' $t$. The Acuna model aggregates all years in the data sample together ($t =$ All, black curve), and in doing so it yields a respectable prediction of $h(t+\Delta t)$ even up to $\Delta t = 6$ years.  However, we find that the model's predictive power depends strongly on the mixing of the age cohorts.  

To demonstrate the model's  dependence on  mixing of career ages, we also show the Acuna model  $R^{2}(t, \Delta t)$  calculated without aggregating data across  all $t$ (colored curves). From this one can clearly see that the $R^{2}(t, \Delta t)$ values calculated for a  fixed $t$ are significantly less than the $R^{2}(\Delta t)$ values calculated by aggregating across  all career ages. This means the model is generally poor at predicting the future success of early career scientists. We also note that artificially large $R^{2}$ values can follow from predictability models which  use cumulative  variables (e.g. $h(t)$ which is non-decreasing), as opposed to incremental variables, such as $\Delta h(t,\Delta t) \equiv h(t+\Delta t)- h(t)$ \cite{SpuriousRegressions, TechnicalPredictabilityPaper}. These limitations are particularly concerning as early career decisions make up a significant portion of cases where quantitative approaches are likely to be applied. 
We further confirmed our observation of much lower $R^{2}$ values in the early career ($t$ up to 5 years) using additional career data for 200 highly cited physicists.

Recent work by A. Mazloumian hints at one of the underlying difficulties of predicting a scientist's future success~\cite{Mazloumian}. By differentiating between citations accrued by papers {\it already  published} at the time of prediction and citations accrued by future papers {\it  published after} the prediction time, it is shown  that regression approaches do a reasonable job predicting future citations to past papers, but are unable to reliably predict {\it future} citations to {\it future} papers. In the context of predicting the future impact of a scientist, this means  that there is not necessarily a correlation between the impact of papers published in the past and the impact of papers published in the future.

Going forward, these approaches and their successors will be increasingly
exploited in real decision making processes. However, at the moment
little is known about the strengths and weaknesses of the state-of-the-art
predictive indicators.
It is open to debate with whom the {\it responsibility} to 
vet current and new quantitative measures lies. But what is clear is that
scientists themselves, particularly young ones, stand to lose the most
should quantitative measures be stretched too far in the realm of career
decisions. As a community it behooves us to engage with the institutions
that seek to exploit quantitative measures of scientific impact in their
decision making process, while maintaining a skepticism backed by
quantitative and rigorous analysis of the specific measures they seek to
employ.

\end{document}